# The Pressure Effects on Electronic Structure of Iron Chalcogenide Superconductors FeSe$_{1-x}$Te$_x$

A. Ciechan,[1,*] M. J. Winiarski,[2,†] and M. Samsel-Czekała[2,‡]

[1]*Institute of Physics, Polish Academy of Sciences*
*al. Lotników 32/46, 02-668 Warsaw, Poland*
[2]*Institute of Low Temperature and Structure Research, Polish Academy of Sciences*
*ul. Okólna 2, 50-422 Wrocław, Poland*

We study the electronic structure of iron-based superconductors $FeSe_{1-x}Te_x$ within the density functional theory. We pay particular attention to the pressure effects on the Fermi surface (FS) topology, which seem to be correlated with a critical superconducting temperature $T_C$ of iron chalcogenides and pnictides. A reduction of the FS nesting between hole and electron cylinders with increasing pressure is observed, which can lead to higher values of $T_C$. The tellurium substitution into selenium sites yields FS changes similar to the pressure effect.

## I. INTRODUCTION

Iron chalcogenides FeSe$_{1-x}$Te$_x$ are members of promising family of Fe-based high temperature superconductors[1]. Non-stoichiometric Fe$_{1+\delta}$Se have been found to be superconducting at 8K[2]. Tellurium substitution into selenium sites raises the critical temperature $T_C$ in FeSe$_{1-x}$Te$_x$ up to 15K for $x = 0.5$[3–5]. Additionally, $T_c$ increases to 37K for FeSe[6–10] and to 26K for FeSe$_{0.5}$Te$_{0.5}$[11,12] under external pressure. On the other hand, the end member FeTe ($x = 1$) is no longer superconducting, but shows antiferromagnetic phase at low temperatures. Disorder has also influence on chalcogenide properties. It can be introduced by excess iron atoms (deficiency of selenium) in Fe$_{1+x}$Se (FeSe$_{1-x}$) layer[13] or by doping with Ni, Co and Cu into Fe sites[14–18]. Recently, ternary compounds A$_x$Fe$_2$Se$_2$ with alkai metal A = K, Rb, Tl, Cs between FeSe layers have been investigated due to promising $T_C > 30K$[19–22].

These compounds containing no arsenic atoms, unlike pnictide superconductors, are particularly important for applications. They are also convenient for theoretical investigations because of simple both chemical compositions and crystal structures.

The main aim of this paper is to examine the pressure effect on electronic structure of FeSe$_{1-x}$Te$_x$ in the normal state within the density functional theory (DFT) calculations. We describe our computational methods and give structural parameters obtained by a geometry optimization under pressure for FeSe and FeSe$_{0.5}$Te$_{0.5}$ compositions in the tetragonal phase of the PbO-type (P4/nmm). The results of band structure calculations corresponding to ambient and higher pressures are presented. We are especially interested in the changes of the Fermi surface (FS) topology under pressure, which is suspected to be correlated with superconducting temperatures of iron chalcogenides and pnictides.

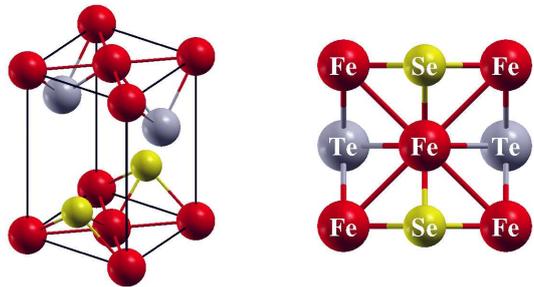

FIG. 1: Schematic tetragonal crystal structure of FeSe$_{1-x}$Te$_x$ ($x = 0.5$) of the PbO-type (space group $P4/nmm$).

## II. COMPUTATIONAL DETAILS

We have studied FeSe$_{1-x}$Te$_x$ superconductors with $x = 0$ and 0.5 displayed in Fig. 1. All calculations were performed in the framework of DFT within the local-density approximation (LDA) of the exchange-correlation potential. We used the based on plane-waves and PAW (Projector Augmented Wave) methods QUANTUM-ESPRESSO code[23] and ABINIT[24] to optimize both lattice parameters and atomic positions in the unit cell. Then we employed the FPLO (full-potential local-orbital) code[25] to calculate all electronic properties of FeSe$_{1-x}$Te$_x$ under external pressure.

The following valence configurations were used in our calculations: $3d^64s^24p^0$, $3d^{10}4s^24p^4$ and $4d^{10}5s^25p^4$ for Fe, Se and Te, respectively. Total energy of considered systems was converged with accuracy to $10^{-4}$Ry for the plane waves energy 50Ry cut-off. The $12 \times 12 \times 12$ (196 points) k-point mesh in the non-equivalent part of the Brilouin zone was sufficient.

The first step of the analysis was geometry relaxation of the FeSe$_{1-x}$Te$_x$ crystal structure in the tetragonal phase, which was performed with 0.05GPa convergence criterion on the pressure ($10^{-3}$Ry/Bohr on forces). The calculated lattice parameters at ambient pressure are pre-sented in Table I. The results of full geometry relaxation differs from both experimental data[7,8,12] and results of calculations limited to optimization of the free $z_{Se/Te}$ parameters[26–28,30].

TABLE I: Experimental and calculated lattice constans $a$, $c$, and free $z_{Se/Te}$ parameters of FeSe and FeSe$_{0.5}$Te$_{0.5}$ at $p = 0$. The bulk moduls $B_0$ obtained from the third-order Birch-Murnaghan equation of state.

|  | $a$[Å] | $c$[Å] | $z_{Se}/z_{Te}$ | $B_0$[GPa] |
| --- | --- | --- | --- | --- |
| FeSe(exp.)[a] | 3.7742 | 5.4545 | 0.266 | 30.7 |
|  | 3.7658 | 5.4988 | 0.266 | 31 |
| FeSe(opt.) | 3.5963 | 5.4310 | 0.256 | 32.9 |
| Fe$_{1.03}$Se$_{0.57}$Te$_{0.43}$(exp.)[a] | 3.8007 | 5.9926 | 0.274 | 36.6 |
| FeSe$_{0.5}$Te$_{0.5}$(exp.)[a] | 3.8003 | 5.9540 | 0.256/0.285 |  |
| FeSe$_{0.5}$Te$_{0.5}$(opt.) | 3.6546 | 5.6847 | 0.238/0.289 | 33.6 |

[a]According to [7,8,12] and [30], respectively.

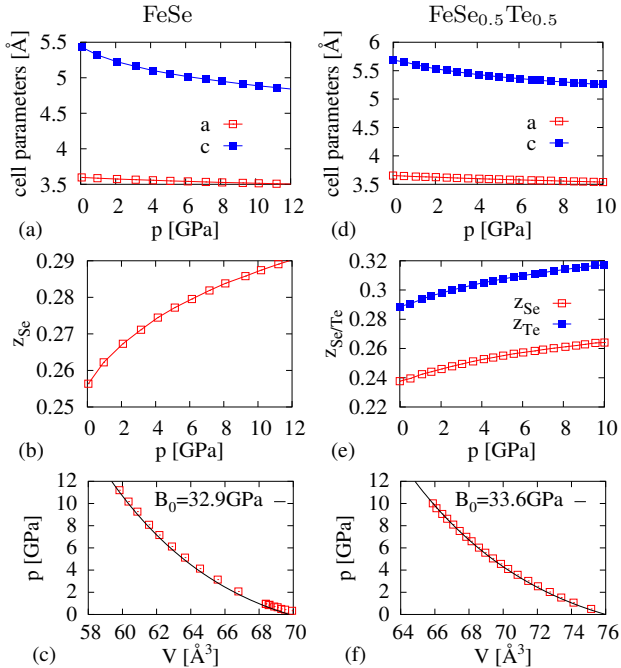

FIG. 2: Calculated pressure evolution of the lattice parameters $a$, $c$ (a and d) as well as the chalcogen atom distance from the iron plane (b and e) for FeSe and FeSe$_{0.5}$Te$_{0.5}$. Their corresponding unit cell-volume changes vs. pressure (squares), fitted to the equation of states (lines), are shown in parts (c and f), respectively.

Figure 2 shows the pressure dependence of the cell parameters and Fe/Se distance from the iron plane in FeSe and FeSe$_{0.5}$Te$_{0.5}$ compounds. We fitted our data with the third order Birch-Murnaghan equation of state[29]:

$$p(V) = \tfrac{3}{2} B_0 \left[ \left(\tfrac{V_0}{V}\right)^{7/3} - \left(\tfrac{V_0}{V}\right)^{5/3} \right] \left[ 1 - \tfrac{3}{4}(4 - B'_0)\left[\left(\tfrac{V_0}{V}\right)^{2/3} - 1\right] \right],$$

where $B_0$ is the bulk modulus, $B'_0$ is its pressure derivative and $V_0$ is the equilibrium unit cell volume. The obtained $B_0$ parameters are in good agreement with the previous experimental data[7,8,12] (Table I).

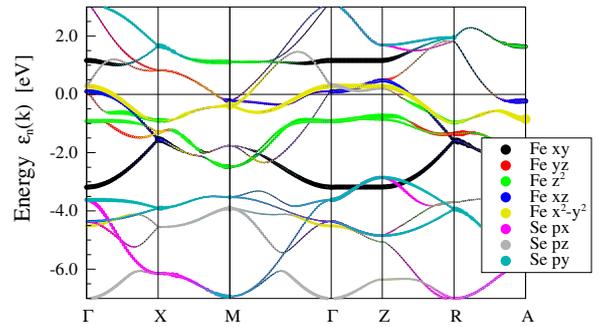

FIG. 3: The electronic band structures of FeSe along high-symmetry lines at $p = 9$GPa.

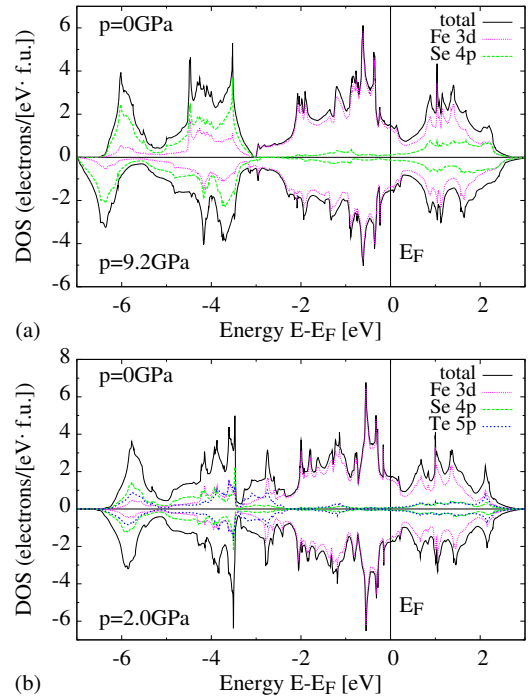

FIG. 4: The total and orbital projected electronic DOS for (a) FeSe and (b) FeSe$_{0.5}$Te$_{0.5}$ at ambient and higher pressures.

## III. ELECTRONIC STRUCTURE

The electronic structure of FeSe$_{1-x}$Te$_x$ near the Fermi energy contains mainly the Fe-3$d$ states[26] as can be seen from the distinct orbital character of bands (Fig. 3 for FeSe at $p \cong 9$GPa) as well as orbital-projected densities of states (DOS) plotted in Fig. 4. However, the bands at the Fermi level ($E_F$) have also contributions from the $p_z$ orbitals, which indicates that Te/Se-$p$ orbitals may play a substantial role in superconductivity as well. A higher DOS at $E_F$ is observed in FeSe$_{0.5}$Te$_{0.5}$ than in FeSe. This suggests a small increase of electronic den-sity with raising tellurium content. In both compounds, the densities at the Fermi energy are slightly enhanced under pressure. Zero pressure values of DOS at $E_F$ are 1.49 eV$^{-1}$ for FeSe and 1.73 eV$^{-1}$ for FeSe$_{0.5}$Te$_{0.5}$. At pressures corresponding to the maximum critical temperatures, the densities are equal to 1.54 eV$^{-1}$ ($p \cong 9$GPA) and 1.77 eV$^{-1}$ ($p \cong 2$GPa), respectively.

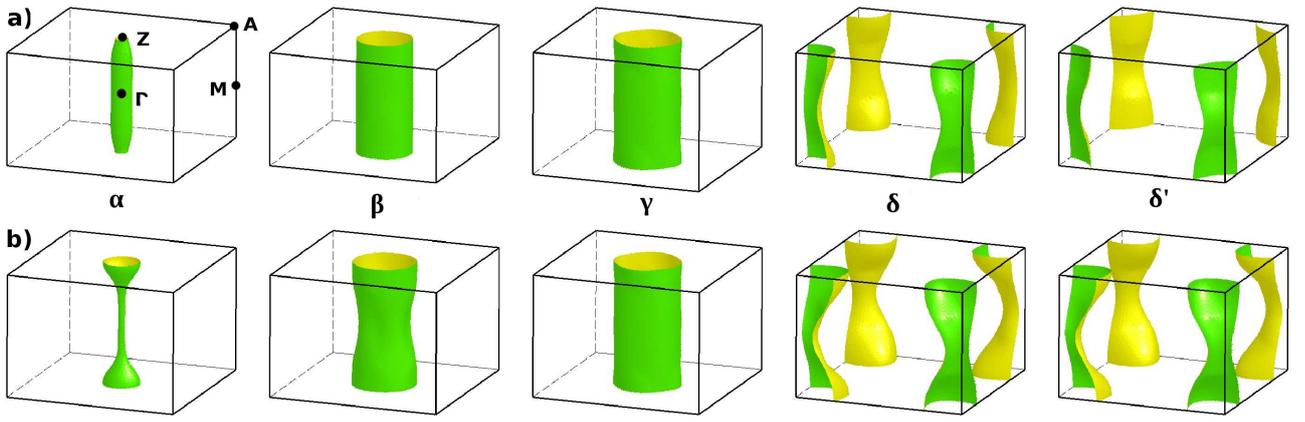

FIG. 5: The Fermi surface sheets of FeSe (drawn separately for each of five bands) at (a) 0 and (b) 9 GPa.

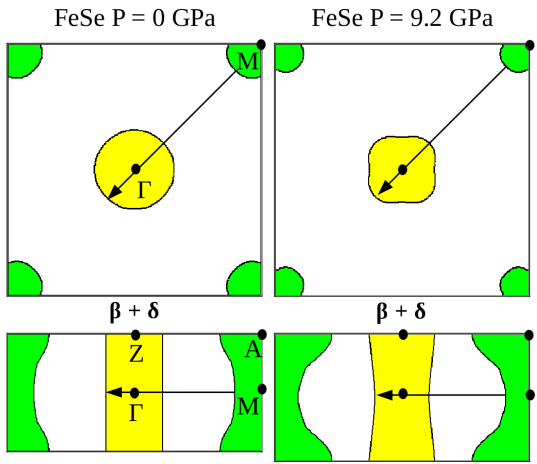

FIG. 6: Changes in the Fermi surface nesting between electron and hole sheets of FeSe under external pressure.

In general, the Fermi surface of FeSe$_{1-x}$Te$_x$ compounds exists in five bands and consists of two electron cylinders $\delta$ and $\delta'$, centred at the M point, and three hole-like sheets around the $\Gamma$ point - two outer cylinders and one inner closed pocket (labelled as $\alpha$, $\beta$, $\gamma$)[26,27]. Figure 5 visualizes all FS sheets for FeSe under ambient and higher pressures. The electron and hole cylinders are separated by the nesting vector close to $\boldsymbol{q} = [\pi, \pi, 0]$ (Fig. 6). Due to this fact, the spin-density waves (SDW) can compete with the superconducting (s-wave-type) pairing. Under external pressure, the cylinders are more corrugated and the FS nesting is suppressed[9].

In turn, the Fermi surface sheets in FeSe$_{0.5}$Te$_{0.5}$ (not shown) have also more 3-dimensional character in comparison with FeSe system and the cylinders yield more imperfect nesting upon increasing pressure. Thus, both tellurium substitution and external pressure have the similar effects on electronic structure of iron chalcogenides in the tetragonal phase.

## IV. CONCLUSIONS

We have studied the effect of external pressure on both crystal and electronic structures of FeSe$_{1-x}$Te$_x$ superconductors. The influence of tellurium content was also investigated. Correlations between critical temperature, lattice parameters and topology of the Fermi surface were observed. The increase of pressure as well as Te substitution raise the values of $z_{Se/Te}$ causing suppression of the Fermi surface nesting.

In FeSe$_{0.5}$Te$_{0.5}$ at $p = 0$, due to the reduced FS nesting, a possible SDW state becomes more unstable than in FeSe. Hence, the superconducting phase can appear at an earlier stage in the former system. In both compounds, the imperfect FS nesting of the corrugated cylinders is enhanced with increasing pressure, which can lead to higher values of $T_c$.

The orbital character of bands and projected DOS confirm that the electronic structure near the Fermi energy consists of Fe-3$d$ electrons being slightly hybridized with chalcogenide 3$p$/4$p$ states. A higher DOS at the Fermi level is observed under pressure, which usually improves superconducting properties.

In previous experiments, critical temperature reaches maximum under finite pressures and then structural phase transitions take place[7,9,12]. Therefore, the observed $T_C$-increase seems to be caused by the pressure-induced changes in the tetragonal phase. When this phase begins to diminish, the trend is reversed ($T_C$ decreases). Ab initio investigations of such effects will be undertaken in the future.

**Acknowledgments**

This work has been supported by the EC project through the FunDMS Advanced Grant of the European Research Council (FP7 'Ideas'), managed by Tomasz Dietl (A.C.) as well as by the National Centre for Science in Cracov under the grant No. N N202 239540 (M.W., M.S-C.). Calculations were partially performed on ICM supercomputers of Warsaw University under the grant G46-13 (A.C) and in Wroclaw Centre for Networking and Supercomputing, grant No.158 (M.W.). The Computing Centre at the Institute of Low Temperature and Structure Research PAS in Wroclaw is acknowledged for the use of the supercomputers and technical support.